\documentclass[twocolumn,prb,superscriptaddress,showpacs,showkeys,floatfix]{revtex4}
\DeclareMathAlphabet{\mathitbf}{OML}{cmm}{b}{it}
\usepackage{graphicx}
\usepackage{dcolumn}
\usepackage{bm}
\usepackage{textcomp}

\begin{document}


\title{Comparative study of the phonons in non-superconducting BaC$_6$ and superconducting CaC$_6$ using inelastic x-ray scattering}


\author{A. C. Walters}   
\email{andrew.walters@esrf.fr}
    \affiliation{ESRF, Polygone Scientifique Louis N\'{e}el, 6 rue Jules Horowitz, 38000 Grenoble, France}
    \affiliation{London Centre for Nanotechnology and Department of Physics and Astronomy, University College London, London WC1E 6BT, United Kingdom}

\author{C. A. Howard}
    \affiliation{London Centre for Nanotechnology and Department of Physics and Astronomy, University College London, London WC1E 6BT, United Kingdom}  
    
\author{M. H. Upton}
 \affiliation{Advanced Photon Source, Argonne National Laboratory, Argonne, IL 60439, USA}
 \affiliation{Condensed Matter Physics and Materials Science Department,Brookhaven National Laboratory, Upton, New York 11973}

\author{M. P. M. Dean}
	\affiliation{Condensed Matter Physics and Materials Science Department,Brookhaven National Laboratory, Upton, New York 11973}
    
\author{A. Alatas}
    \affiliation{Advanced Photon Source, Argonne National Laboratory, Argonne, IL 60439, USA}

\author{B. M. Leu}
 \affiliation{Advanced Photon Source, Argonne National Laboratory, Argonne, IL 60439, USA}
 
\author{M. Ellerby}
\affiliation{London Centre for Nanotechnology and Department of Physics and Astronomy, University College London, London WC1E 6BT, United Kingdom}

\author{D. F. McMorrow}
\affiliation{London Centre for Nanotechnology and Department of Physics and Astronomy, University College London, London WC1E 6BT, United Kingdom}

\author{J. P. Hill}
  \affiliation{Condensed Matter Physics and Materials Science Department,Brookhaven National Laboratory, Upton, New York 11973}

\author{M. Calandra}
    \affiliation{Universit\'{e} Pierre et Marie Curie, 4 Place Jussieu-case postale 115, 72252 Paris, CEDEX 05, France}
 
\author{F. Mauri}
    \affiliation{Universit\'{e} Pierre et Marie Curie, 4 Place Jussieu-case postale 115, 72252 Paris, CEDEX 05, France}
    

\date{\today}

\begin{abstract}

The low energy phonons of two different graphite intercalation compounds (GICs) have been measured as a function of temperature using inelastic x-ray scattering (IXS). In the case of the non-superconductor BaC$_6$, the phonons observed are significantly higher (up to 20\%) in energy than those predicted by theory, in contrast to the reasonable agreement found in superconducting CaC$_6$. Additional IXS intensity is observed below 15 meV in both BaC$_6$ and CaC$_6$. It has been previously suggested that this additional inelastic intensity may arise from defect or vacancy modes not predicted by theory (d'Astuto et al, Phys. Rev. B \textbf{81} 104519 (2010)). Here it is shown that this additional intensity can arise directly from the polycrystalline nature of the available samples. Our results show that future theoretical work is required to understand the relationship between the crystal structure, the phonons and the superconductivity in GICs. 

\end{abstract}

                             

\maketitle


\section{Introduction\label{intro}}

Since the discovery of superconductivity in YbC$_6$ ($T_c = 6.5$ K) and CaC$_6$ ($T_c = 11.5$ K) at temperatures over an order of magnitude higher than previously found in graphite intercalation compounds (GICs) \cite{Weller_NP_2005,Emery_PRL_2005}, the properties of this family of GICs have been extensively studied by a variety of different experimental techniques \cite{Cubitt_PRB_2007,Gauzzi_PRB_2008,Kim_PRL_2007,Hinks_PRB_2007,Gonnelli_PRL_2008,Valla_PRL_2009}. Although initially an exotic superconducting mechanism was proposed involving acoustic plasmons \cite{Csanyi_NP_2005}, subsequent density functional theory (DFT) studies described the superconductivity via a more orthodox electron-phonon (e-ph) coupling mechanism with s-wave symmetry \cite{Calandra_PRL_2005,Calandra_PRB_2006,Sanna_PRB_2007,Calandra_PRB_2010}. These DFT descriptions predict that the e-ph coupling is approximately equal for phonons associated with vibration of the carbon atoms and for phonons associated with movement of the intercalant calcium. 

At present there are experimental studies in the literature which give conflicting viewpoints concerning the nature of the e-ph coupling in GICs. A large Ca isotope effect ($\alpha$(Ca) $\sim$ 0.5) has been measured in CaC$_6$ \cite{Hinks_PRB_2007}, which, if viewed within the BCS description of superconductivity, suggests that only the phonons due to the vibration of calcium are involved in the electron pairing. In contrast, angle-resolved photoemission spectroscopy (ARPES) measurements on CaC$_6$ \cite{Valla_PRL_2009} have found that the e-ph coupling to graphite-like high-energy phonons is so strong that it can explain the superconducting transition temperature alone, without any additional coupling to calcium phonons. These discrepancies point to the need for a detailed study of the phonons in GICs, to both test the DFT description and to look for direct evidence for e-ph coupling involving specific phonons. Moreover, phonon studies in graphitic systems in general are important as the electron-phonon interactions in these systems are under much scrutiny \cite{Pisana_NM_2007,Lazzeri_PRL_2006,Dean_PRB_2010a}.

Superconductivity in GICs is directly linked to the graphite layer separation, with the superconducting transition temperature $T_c$ increasing as the graphite layer separation $d$ is reduced. This trend is supported both by the observed values of $T_c$ in a variety of GICs \cite{Kim_PRL_2007} and by measurements of $T_c$ as a function of pressure in CaC$_6$ \cite{Gauzzi_PRL_2007} and YbC$_6$ \cite{Smith_PRB_2006}. Indeed in BaC$_6$ the graphite layer separation is so large that superconductivity appears to be suppressed entirely, with no superconducting transition observed down to 0.080 K \cite{Nakamae_SSC_2008}. It is therefore instructive to study the superconductivity in GICs by studying GICs with different intercalants, since by changing the intercalant one changes $d$ and therefore tunes $T_c$. 

In this paper we present the low energy phonon dispersions in non-superconducting BaC$_6$ and superconducting CaC$_6$ as measured using inelastic x-ray scattering (IXS). These data represent the first momentum-resolved phonon measurements on BaC$_6$. We find a substantial discrepancy between experiment and theory in the phonon energies of BaC$_6$, in contrast to the good agreement in the case of CaC$_6$. Like many other layered materials, GICs are difficult to synthesise as high-quality single crystals. We demonstrate here that the details of the preferred orientation (texture) of the crystallites in these GIC samples can lead to the observation of phonons which, because of their polarization, are theoretically forbidden to be observed. Our work highlights the importance of accounting for these effects in phonon studies of textured polycrystalline samples and calls for the need for single crystal GIC samples to address the role of the electron-phonon coupling in these materials.

The low energy phonons in CaC$_6$ have been measured previously \cite{Upton_PRB_2007} and found to be in good overall agreement with the published DFT calculations. Subsequent to this work two momentum-resolved phonon studies have been made on CaC$_6$, an inelastic neutron scattering (INS) study, which concentrates primarily on the high energy graphite-like phonon modes \cite{Dean_PRB_2010b}, and a study of the low energy phonons in CaC$_6$ performed using both IXS and INS \cite{dAstuto_PRB_2010}. While in Ref. \cite{Dean_PRB_2010b} good agreement was found with the calculated DFT phonon dispersions by taking account of the polycrystalline nature of the sample, in Ref. \cite{dAstuto_PRB_2010} it was suggested that an additional phonon mode exists in CaC$_6$ of uncertain origin. Here we propose an alternative explanation: that the additional IXS intensity arises from the weak crystallographic texture in the polycrystalline GIC samples.

\section{Experimental methods\label{methods}}

\begin{figure}
	\begin{center}
		\includegraphics[width=7.3cm]{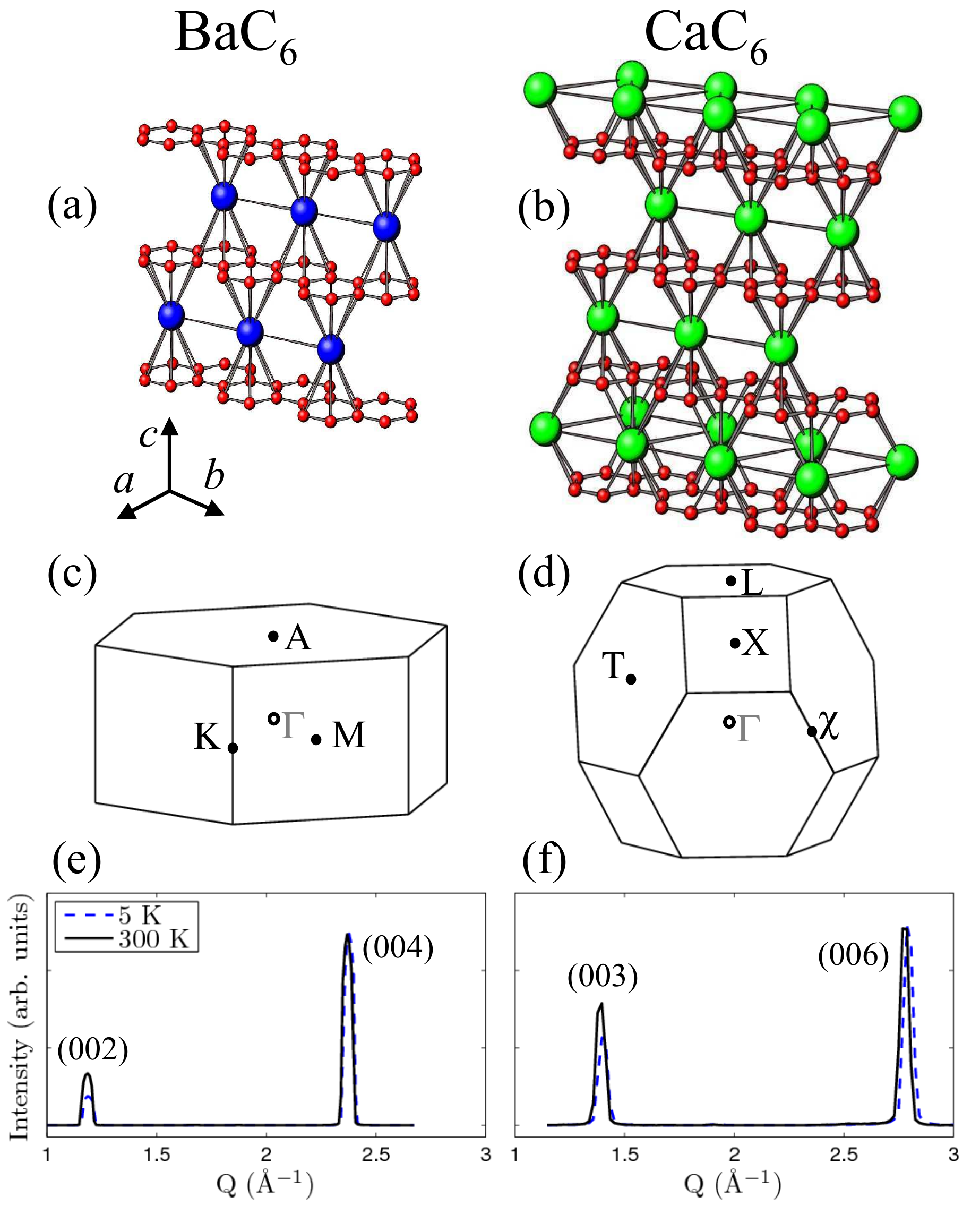}
		\caption{The structural properties of BaC$_6$ and CaC$_6$: (a)-(b) Crystal structure, (c)-(d) First Brillouin zone with symmetry points, (e)-(f) Diffraction on BaC$_6$ and CaC$_6$ measured in (00$l$) direction at 5 K (dashed blue line) and at 300 K (solid black line) at 3-ID. The symmetry points shown are at the edge of the Brillouin zone apart from the $\Gamma$ point, which is at the zone centre. There is practically no signal from impurities in the (00$l$) diffraction from either sample. \label{figure_structural}}
	\end{center}
\end{figure}

The GIC samples were made using ZYA grade highly orientated pyrolytic graphite (HOPG) platelets purchased from GE Advanced Ceramics. The BaC$_6$ sample was made using the vapour transport method \cite{Dresselhaus_AiP_2002}. An HOPG platelet was outgassed at 500$^{\circ}$C and then sealed into a quartz tube along with the barium metal (purity $>$ 99.99\%) under high vacuum ($<10^{-6}$ mbar). The tube was then heated to 490$^{\circ}$C and maintained at this temperature for 4 weeks. The CaC$_6$ samples used were prepared by immersing a HOPG platelet in a Li-Ca alloy for 10 days, as described elsewhere \cite{Pruvost_C_2004}. CaC$_6$ samples made from the same batch were found to have a sharp superconducting transition at 11.5 K from magnetic susceptibility measurements. The very high purity of the samples ($>$ 99\% pure in both cases) can be seen in the (00$l$) diffraction shown in Figure \ref{figure_structural} (e) and \ref{figure_structural} (f), where there are no visible Bragg peaks from any impurities.

The starting graphite (HOPG) is composed of small crystallites ($\approx$ 1 $\mu$m) which have a strong preferred orientation (strong texture) perpendicular to the graphene planes (out-of-plane), giving a (00$l$) mosaic with full-width half-maximum (FWHM) as low as 0.2$^{\circ}$. However within the graphene planes (in-plane), the crystallites are orientated randomly \cite{Dresselhaus_AiP_2002}. After intercalation the crystallites are still oriented randomly in-plane, but out-of-plane the orientation of the crystallites is more random, with the GICs studied here having (00$l$) mosaics of 5$^{\circ}$. This means that the texture is weaker in the GIC samples than in HOPG, since a weaker texture means that the samples are more like a perfect powders, which have zero texture.

CaC$_6$ has the structure $R\bar{3}m$, which can be described using a rhombohedral or a hexagonal basis \cite{Emery_PRL_2005}. The calcium atoms are arranged in three different ways in different intercalant layers (called A$\alpha$A$\beta$A$\gamma$ stacking, where the Roman letters define graphite layers and the Greek letters intercalant layers). The unit cell of CaC$_6$ in the hexagonal basis is shown in Figure \ref{figure_structural}(b), and the shape of the first Brillouin zone is shown in Figure \ref{figure_structural}(d). Here we define the reciprocal lattice directions using the hexagonal basis, meaning that the out-of-plane direction is the (00$l$) direction. This convention aids comparison with the BaC$_6$ data, as BaC$_6$ has the space group $P6_{3}/mmc$, which is normally described within the hexagonal basis. The stacking in BaC$_6$ is A$\alpha$A$\beta$, as shown in Figure \ref{figure_structural}(a). The first Brillouin zone of BaC$_6$ is presented in Figure \ref{figure_structural}(c). The lattice parameters of CaC$_6$ are $a=4.333(2)$ \AA $ $ and $c=13.572(2)$ \AA \cite{Emery_JoSSC_2005}, giving a graphite layer separation $d$ of $4.524(1)$ \AA, and in BaC$_6$ the lattice parameters are $a=4.302(6)$ \AA $ $ and $c=10.50(4)$ \AA $ $ with $d=5.25(2)$ \AA\cite{Guerard_C_1980}.

The IXS measurements were performed at 3-ID at the Advanced Photon Source, Argonne National Laboratory, with an incident x-ray energy of 21.657 keV\cite{Toellner_JoSR_2006}, providing an energy resolution between 2.2 and 2.4 meV, depending on the specific analyzer. The spectrometer has four analyzers, allowing energy scans to be done at different momentum transfers simultaneously. The momentum resolution in all cases was 0.072 \AA$^{-1}$ in the scattering plane and 0.183 \AA$^{-1}$ perpendicular to it. The phonon peaks were fitted with pseudo-Voigt functions, which were appropriately scaled by the Bose factor. The phonon dispersions were produced by plotting the peak positions as a function of the phonon wavevector $\mathitbf{q}$, defined within the equation $\mathitbf{Q}=\mathitbf{G}+\mathitbf{q}$, where $\mathitbf{G}$ is the nearest reciprocal lattice vector and $\mathitbf{Q}$ the momentum transfer.

In order to model the effect of the crystallographic texture in the GIC samples on the phonon spectra, simulations of the IXS data were produced by summing hundreds of simulated IXS intensities, each of which was performed at a specific momentum transfer. The crystallographic texture was described by performing simulations over a volume in reciprocal space expressed in spherical polar coordinates ($|\mathitbf{Q}|$,$\theta$,$\phi$), where $\theta$ has its rotation axis out-of-plane. For the 2D powder simulations, the momentum transfers were selected using Lorentzian sampling of $\phi$ with a FWHM of 5$^{\circ}$ and allowing $\theta$ to take any value. For the 3D powder simulations both $\phi$ and $\theta$ were allowed to take any value. The summed IXS spectra were then convolved with the momentum and energy resolution of the IXS spectrometer. This method was also used in our recent INS study of CaC$_6$ \cite{Dean_PRB_2010b}. 

The two-dimensionality of these GICs, together with the significant difference in mass between the intercalant and carbon, means that the phonon modes can be separated, to a good approximation, into four groups: I$_{xy}$, I$_{z}$, C$_{xy}$ and C$_{z}$, where I$_{xy}$ describes phonon modes purely due to vibrations of intercalant atoms in-plane, I$_{z}$ the intercalant phonons out-of-plane, and C$_{xy(z)}$ the equivalent carbon in-plane (out-of-plane) phonons.

\section{Inelastic x-ray scattering measurements\label{ixs}}

\subsection{Phonons in BaC$_6$\label{bac6}}

\begin{figure}
	\begin{center}
		\includegraphics[width=7.0cm]{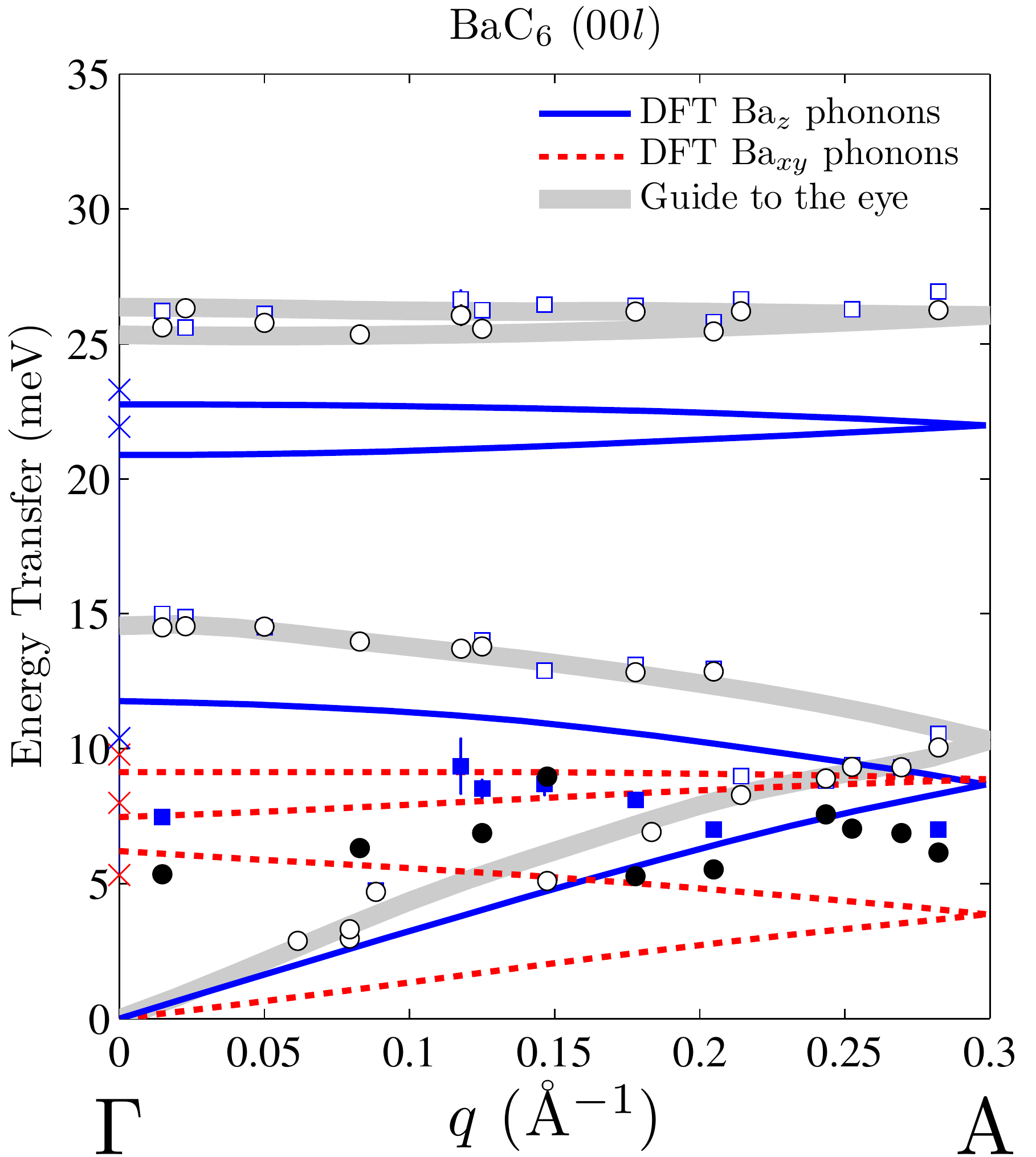}
		\caption{BaC$_6$ (00$l$) phonon dispersion measured at 5 K (squares) and 300 K (circles). Empty symbols denote Ba$_z$ phonon intensity, full symbols label Ba$_{xy}$ phonon intensity. The Ba$_{xy}$ phonons should not be visible in this scattering geometry (see text). The theoretical dispersion of the Ba$_{z}$ and Ba$_{xy}$ phonons are plotted with solid lines and dashed lines respectively \cite{Calandra_PRB_2006}. A guide to the eye is plotted showing the dispersion of the Ba$_z$ phonons at 300 K (thick solid line). The symmetry point A is located at $q= 0.299$ \AA$^{-1}$. The crosses mark the phonon energies calculated at $\Gamma$ using the experimental lattice parameters and space group of BaC$_6$.  \label{figure_BaC6_dispersion}}
	\end{center}
\end{figure}

The (00$l$) phonon dispersions of BaC$_6$ measured at both 5 K and 300 K are plotted in Figure \ref{figure_BaC6_dispersion}, together with calculated dispersions \cite{Calandra_PRB_2006}. The measured Ba$_z$ phonon branches are significantly higher in energy than the theoretical values: near the edge of the Brillouin zone, at symmetry point A, the measured phonon energy for the optic Ba$_{z}$ mode is almost $20\%$ larger than that predicted. In addition, in all of the measured IXS energy scans on BaC$_6$ there is additional intensity observed below 10 meV, plotted with full symbols, which is not predicted by theory.

So what is the origin of this discrepancy between experiment and theory in BaC$_6$? We consider a number of possibilities. The DFT calculations were performed using a different structure to the structure experimentally determined, which may have affected their results. In addition, the charge transfer from the Ba atoms to the graphene planes may be inaccurately predicted by theory. Finally, the calculations do not account for the polycrystallinity of the real samples, so part of the disagreement may be due to their crystallographic texture. 

The published calculations for BaC$_6$ use lattice parameters equivalent to $a= 4.350$ \AA $ $ and $c=10.40$ \AA \cite{Calandra_PRB_2006} which are significantly different to the values found via x-ray diffraction ($a=4.302(6)$ \AA $ $ and $c=10.50(4)$ \AA) \cite{Guerard_C_1980}. In addition, the space group of CaC$_6$ ($R$\={3}$m$) is used for BaC$_6$ in the calculation, rather than the experimentally found $P6_{3}/mmc$. To determine whether the observed discrepancy arises because of these structural differences, we performed an additional phonon calculation at the $\Gamma$ point using the experimental structure and lattice parameters, as shown in Figure \ref{figure_BaC6_dispersion} (crosses). This calculation gives phonon energies in approximate agreement with the previous calculation, showing that the calculations are largely insensitive to small changes in both the lattice parameters and the space group. Therefore the incorrect structure used in the initial calculations can be eliminated as a cause of the discrepancies.

Another possibility is that the charge transfer from the Ba atoms to the graphite has been underestimated theoretically. If there is less charge than predicted in the graphitic $\pi^*$ bands, then the bonds should be stronger than predicted, since filling the anti-bonding $\pi^*$ band destablizes the bonds. Stronger bonds lead directly to higher phonon energies. The effect of charge transfer on phonons in GICs has been predicted theoretically \cite{Boeri_PRB_2007} and observed using Raman scattering \cite{Dean_PRB_2010a}. This would explain why the phonons are higher in energy than predicted, but cannot explain why additional intensity is observed below 10 meV.

The additional phonon intensity can be explained as a result of the crystallographic texture of the GIC samples. Here we argue that this signal arises from the excitation of I$_{xy}$ phonons over a large volume in reciprocal space due to the weak crystallographic texture. For an ideal single crystal, and perfect instrumental resolution, the I$_{xy}$ phonons are disallowed (their intensity is zero) when $\mathitbf{Q}$ is completely out-of-plane due to the $\mathitbf{Q}\cdot\mathitbf{e}(\mathitbf{q})$ term in the IXS phonon cross-section, where $\mathitbf{e(q)}$ is the eigenvector of the phonon with wavevector $\mathitbf{q}$ \cite{Burkel_RoPiP_2000}. However in the case of the polycrystalline GIC samples studied, the weak crystallographic texture provides an explanation for the observation of the I$_{xy}$ phonons. Even if the nominal momentum $\mathitbf{Q}$ is entirely out-of-plane, the large reciprocal space volume integrated over in each measurement due to the weak preferred orientation of the crystallites will include many values of $\mathitbf{Q}$ which have a significant component in-plane. This effect is discussed further in Section \ref{disc}.

\subsection{Phonons in CaC$_6$\label{cac6}}

\begin{figure}
	\begin{center}
		\includegraphics[width=7.0cm]{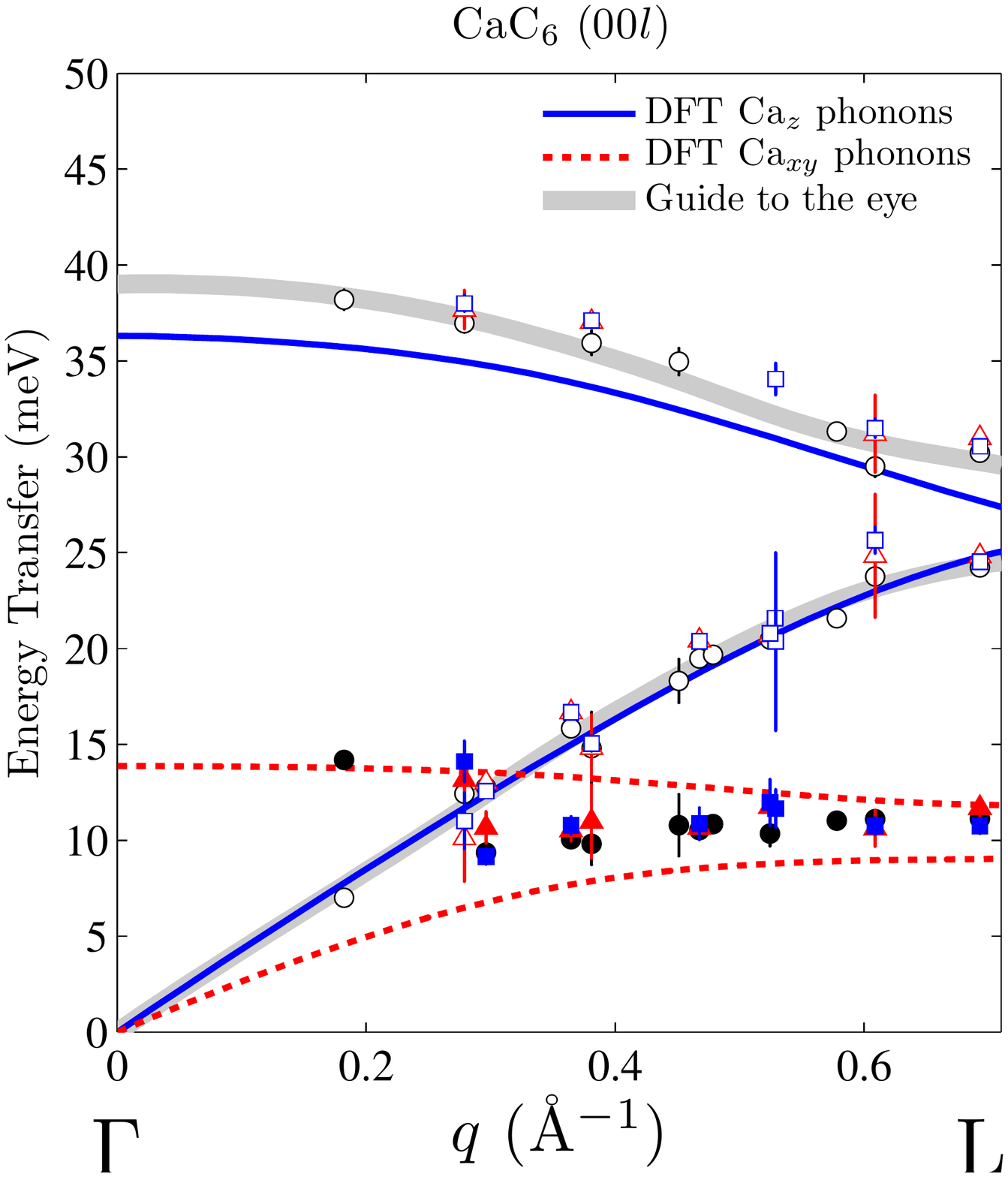}
		\caption{CaC$_6$ (00$l$) phonon dispersion measured at 5 K (squares), 50 K (triangles) and 300 K (circles). Empty symbols denote Ca$_z$ phonon intensity, full symbols label Ca$_{xy}$ phonon intensity. The Ca$_{xy}$ phonons should not be visible in this scattering geometry (see text). The theoretical dispersion of the Ca$_{z}$ and Ca$_{xy}$ phonons are plotted with solid lines and dashed lines respectively \cite{Calandra_PRL_2005}. A guide to the eye is plotted showing the dispersion of the experimental Ca$_z$ phonons at 300 K (thick solid line). The symmetry point L is located at $q= 0.694$ \AA$^{-1}$. \label{figure_CaC6_dispersion}}
	\end{center}
\end{figure}

Figure \ref{figure_CaC6_dispersion} presents the (00$l$) phonon dispersion in CaC$_6$ at 5 K, 50 K and 300 K. A subset of the CaC$_6$ data has already been published \cite{Upton_PRB_2007}, but the scope of the data presented here is much more extensive. The additional phonon intensity below 15 meV again results from the large mosaic of the sample and is discussed at length in Section \ref{disc}. The phonon dispersions calculated using DFT are plotted on the same figure \cite{Calandra_PRL_2005}. The two Ca$_z$ modes are well described by the DFT calculations over the whole $Q$ range sampled, especially in the case of the acoustic Ca$_{z}$ mode. The higher energy mode which disperses between 30 and 40 meV (the optic Ca$_{z}$ mode) is about 2 meV higher in energy than predicted, but the character of the dispersion is reasonably well described. 

In both CaC$_6$ and BaC$_6$ the energies of the I$_z$ modes are slightly hardened ($<$ 1 meV) upon cooling from 300 K to 5 K, but there is no observable difference in CaC$_6$ between the data measured above and below $T_c$. The small temperature dependence most likely results from the reduction in the $c$ lattice parameter upon cooling, visible in the diffraction presented in Figure \ref{figure_structural}(e) and \ref{figure_structural}(f).

\section{Modelling the crystallographic texture in B\MakeLowercase{a}C$_6$ and C\MakeLowercase{a}C$_6$ \label{disc}}

\begin{figure}
	\begin{center}
		\includegraphics[width=8.3cm]{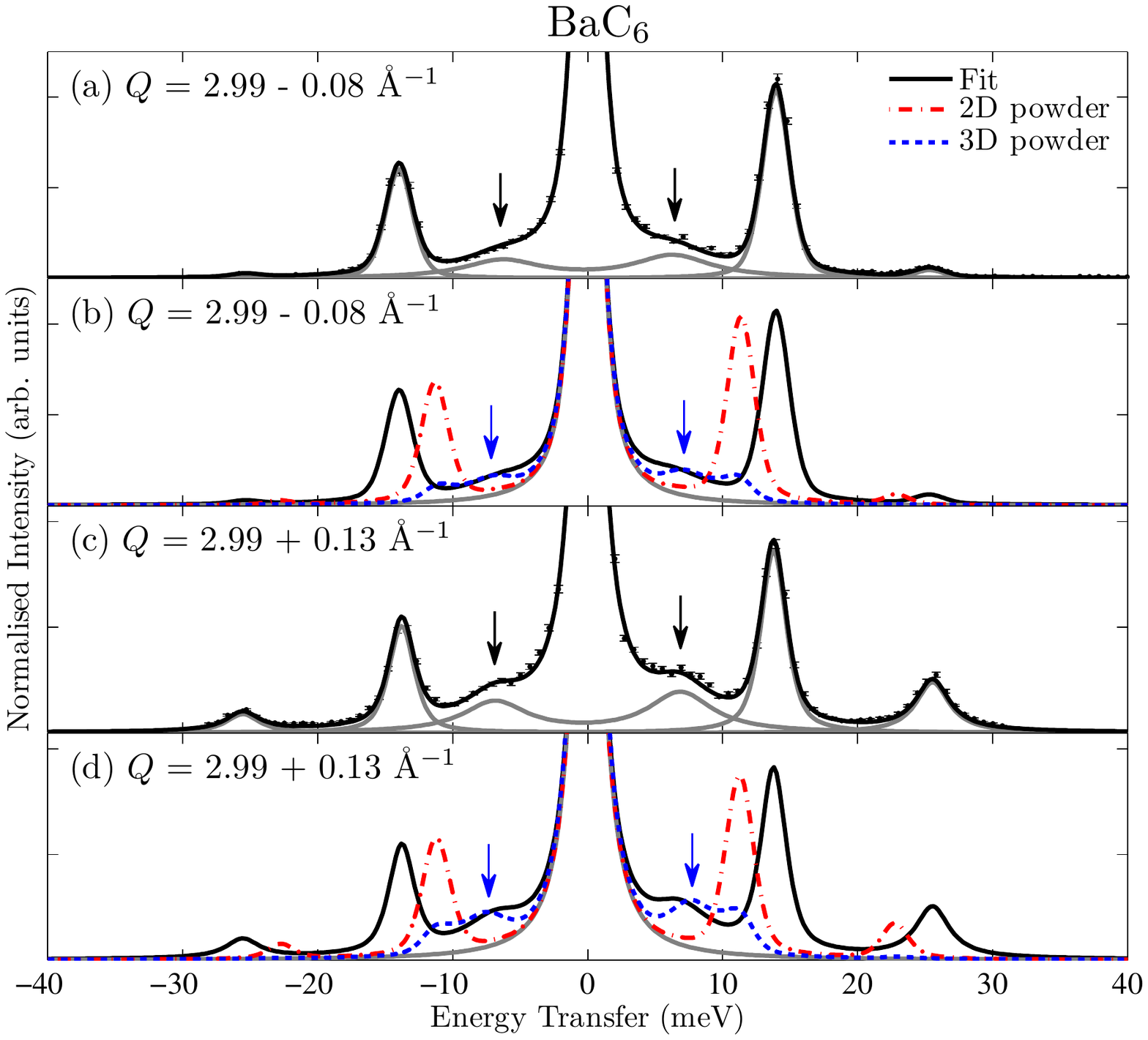}
		\caption{(a) (00$l$) IXS data at $Q$ = 2.91 \AA$^{-1}$ on BaC$_6$ at 300 K, plotted with the data fit and the decomposed fitted phonon peaks. (b) Fit to the data at $Q$ = 2.91 \AA$^{-1}$, plotted with the fitted elastic line and against IXS simulations based on \emph{ab-initio} calculations \cite{Calandra_PRB_2006} which use the 2D powder model and the 3D powder model described in the text. (c) \& (d) Identical to (a) \& (b) but for $Q$ = 3.12 \AA$^{-1}$. In each panel the experimental (simulated) IXS intensity due to Ba$_{xy}$ phonons is marked with black (blue) arrows.  \label{figure_BaC6_rawdata_forpaper}}
	\end{center}
\end{figure}

\begin{figure}
	\begin{center}
		\includegraphics[width=8.3cm]{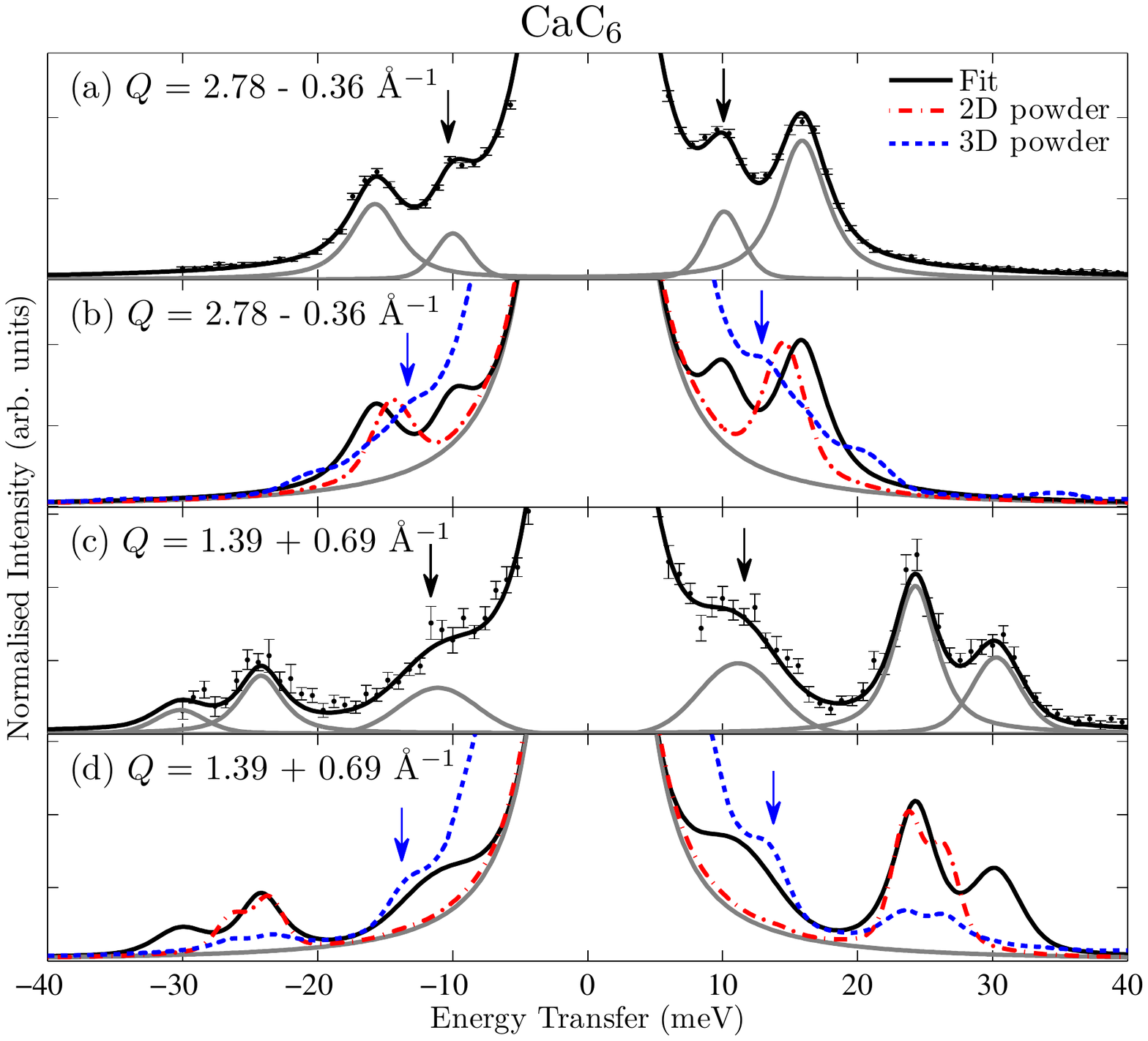}
				\caption{(a) (00$l$) IXS data at $Q$ = 2.42 \AA$^{-1}$ on CaC$_6$ at 300 K, plotted with the data fit and the decomposed fitted phonon peaks. (b) Fit to the data at $Q$ = 2.42 \AA$^{-1}$, plotted with the fitted elastic line and against IXS simulations based on \emph{ab-initio} calculations \cite{Calandra_PRL_2005} which use the 2D powder model and the 3D powder model described in the text. (c) \& (d) Identical to (a) \& (b) but for $Q$ = 2.08 \AA$^{-1}$. In each panel the experimental (simulated) IXS intensity due to Ca$_{xy}$ phonons is marked with black (blue) arrows.  \label{figure_CaC6_rawdata_forpaper}}
	\end{center}
\end{figure}

In Figures \ref{figure_BaC6_rawdata_forpaper} and \ref{figure_CaC6_rawdata_forpaper} a selection of phonon spectra measured at 300 K in the (00$l$) direction in BaC$_6$ and CaC$_6$ are presented. In both cases two phonons are observed in the raw IXS data (the acoustic and optic I$_{z}$ branches), as well as additional IXS intensity at low energies. In each of the panels two IXS simulations are plotted. The first simulation models the sample as a 2D powder: that is, the crystallites have no preferred orientation in-plane, but have a Lorentzian mosaic of FWHM = 5$^{\circ}$ out-of-plane, consistent with our x-ray diffraction. The second simulation models the sample with no preferred orientation (no texture): i.e. as a 3D powder. Both simulations are added to the experimental elastic intensity.

In the case of BaC$_6$, plotted in Figure \ref{figure_BaC6_rawdata_forpaper}, the overall agreement is better with the 2D powder model than with the 3D powder model, especially in the ratio between the acoustic and optic Ba$_z$ phonons. However in the 2D powder model the IXS intensity due to the Ba$_{xy}$ phonons is much smaller than measured. The additional features predicted by the 3D powder model are similar to those observed, but there the intensities of the Ba$_z$ phonons are underestimated. These observations suggest that perhaps the crystallographic texture of the BaC$_6$ samples is weaker than expected from our x-ray diffraction, lying somewhere between the 2D and 3D powder models.

In the CaC$_6$ data in Figure \ref{figure_CaC6_rawdata_forpaper}, the low energy IXS features look qualitatively similar to the Ca$_{xy}$ features predicted by the 3D powder model. However the 3D powder description of CaC$_6$ is once again not a satisfactory description, as the intensity of the Ca$_z$ phonons is underestimated in most cases and the IXS intensity at low energies is significantly overestimated \cite{BaC6paper_comment1}. The measured CaC$_6$ data appear to lie somewhere between the 2D and 3D powder simulations, similar to the BaC$_6$ data, though the CaC$_6$ appears to be more like the 3D powder than BaC$_6$. This suggests that the distribution of intercalant atoms in CaC$_6$ is rather random, which is consistent with the behavior of CaC$_6$ under pressure as studied with x-ray diffraction, where the Ca atoms are found to be very mobile \cite{Gauzzi_PRB_2008}. More theoretical and experimental work is required in order to understand the complicated crystallographic texture of GICs, with more extensive x-ray diffraction being a natural starting point.

A recent IXS and INS study by d'Astuto et al. \cite{dAstuto_PRB_2010} on CaC$_6$ suggested that the additional inelastic intensity was due to an interaction with the acoustic Ca$_z$ mode, causing an avoided crossing, or anti-crossing, which is seen as a splitting in the acoustic Ca$_{z}$ mode. Their work was supplemented by INS data, which allowed them to more easily access momenta nearer to the $\Gamma$ point. The study concluded that the additional inelastic intensity could be due to a defect or vacancy mode. Although we cannot exclude this hypothesis, our simulations suggest that if the orientation of the crystallites in these samples is more random than previously thought, the weak crystallographic texture can account for the anomalous features without recourse to such a mode. 

\section{Conclusions\label{conc}}

To summarise, the dispersions of the low-energy phonons in BaC$_6$ and CaC$_6$ have been measured  as a function of temperature using inelastic x-ray scattering. In BaC$_6$ the experimental and DFT-calculated phonon dispersions \cite{Calandra_PRB_2006} disagree, with measured phonon energies up to 20\% higher than predicted. We suggest that this large discrepancy may result from an underestimation of the charge transfer from the Ba atoms to the graphite sheets in the theory. Our work motivates further study on BaC$_6$ in order to examine the underlying reasons for this disagreement. In contrast, reasonable agreement with theory is found in CaC$_6$ for the Ca$_z$ phonons. This consistency between theory and experiment provides indirect supporting evidence for the DFT description \cite{Calandra_PRL_2005} of the superconductivity in CaC$_6$. 

No signatures of electron-phonon coupling are observed in the phonon dispersions or the phonon widths in either non-superconducting BaC$_6$ or superconducting CaC$_6$, despite the drastically different superconducting transitions of these related compounds. In both BaC$_6$ and CaC$_6$ there is a small ($<$ 1 meV) hardening of the I$_z$ phonons as the temperature is decreased, but this is likely due to a reduction in the $c$ lattice parameter and is unaffected by the presence of superconductivity in CaC$_6$ below 11.5 K. The largest source of phonon broadening experimentally is very likely the weak crystallographic texture inherent in the GIC samples.

Finally, the IXS simulations presented here show that weak crystallographic texture in polycrystalline GIC samples may lead to additional inelastic intensity from I$_{xy}$ phonons. Such additional intensity has been observed recently in CaC$_6$ \cite{Upton_PRB_2007,dAstuto_PRB_2010} and YbC$_6$ \cite{Upton_PRB_2010}, but also in older INS studies on RbC$_{24}$\cite{Funahashi_P_1983} and KC$_{24}$\cite{Magerl_SM_1983}. This work provides a timely reminder that the crystallographic texture inherent in many graphitic systems may give rise to unexpected experimental effects.

\begin{acknowledgments}
We would like to thank Matteo d'Astuto and Michael Krisch for illuminating discussions. A. C. W. would like to thank Ian Wood and Richard Thanki for their assistance and the EPSRC and STFC for funding. Calculations were performed at the IDRIS supercomputing center (project 081202). The work at Brookhaven is supported in part by the US DOE under contract No. DEAC02-98CH10886 and in part by the Center for Emergent Superconductivity, an Energy Frontier Research Center funded by the US DOE, Office of Basic Energy Sciences. The work at the Advanced Photon Source was supported by the US DOE, Office of Basic Energy Sciences, under contract No. DE-AC02-06CH11357.
\end{acknowledgments}

\bibliography{CaC6_BaC6_final}

\begin{thebibliography}{33}
\bibitem[{BaC()}]{BaC6paper_comment1}
\bibinfo{howpublished}{The large IXS intensity at low energy transfer in the
  CaC$_6$ `3D powder' simulations arises from the low energy phonons predicted
  around the X point in the Brillouin zone (shown in Figure
  \ref{figure_structural}(d)), which are intense because the IXS cross-section
  is inversely proportional to energy transfer.}
\end{thebibliography}

\end{document}